\documentclass[twocolumn]{aastex7}

\usepackage[caption=false]{subfig}
\usepackage{mathtools,amssymb}

\newcommand{\vsh}{v_{\rm sh}}

\newcommand{\w}[1]{v_{\rm A,#1}}

\newcommand{\Emax}{E_{\rm max}}

\submitjournal{ApJ}

\shorttitle{Shock Acceleration in the ICM}
\begin{document}

\title{Shock Acceleration in the Intracluster Medium: Implications of Micromirror Confinement}

\begin{abstract}
    Merging galaxy clusters exhibit strong observational evidence for efficient particle acceleration in the intracluster medium (ICM), particularly in the form of synchrotron-emitting radio relics and halos. Cosmic ray (CR) electrons are likely accelerated (or re-accelerated) at merger and accretion shocks via diffusive shock acceleration (DSA). However, in the presence of the large diffusion coefficients one would naively expect in the rarefied, relatively unmagnetized ICM, this acceleration--in particular, the maximum proton energy ($E_{\rm max}$)--is limited by long acceleration times. On the other hand, recent work on CR transport suggests that the diffusion coefficient can be suppressed in ICM-like environments. In this picture, deviations from local thermodynamic equilibrium can trigger the mirror instability, creating plasma-scale magnetic structures, or ``micromirrors," that efficiently scatter CRs. In this paper, we investigate the implications of micromirror confinement for shock acceleration in the ICM. We demonstrate that micromirrors enforce a minimum value of $E_{\rm max} \gtrsim 100$ GeV that does not rely on CR-driven magnetic field amplification. We also discuss micromirror confinement in the context of cosmological simulations and $\gamma$-ray observations, and present a simulation of a Coma-like merging cluster that self-consistently includes CR acceleration at shocks, with an effective diffusion coefficient set by micromirrors. We show that the introduction of micromirrors yields simulated galaxy clusters that remain consistent with $\gamma$-ray observations. 
\end{abstract}

\correspondingauthor{Rebecca Diesing}
\affiliation{School of Natural Sciences, Institute for Advanced Study, Princeton, NJ 08540, USA}
\affiliation{Department of Physics and Columbia Astrophysics Laboratory, Columbia University, New York, NY 10027, USA}
\email{rrdiesing@ias.edu}

\author[0000-0002-6679-0012]{Rebecca Diesing}
\affiliation{School of Natural Sciences, Institute for Advanced Study, Princeton, NJ 08540, USA}
\affiliation{Department of Physics and Columbia Astrophysics Laboratory, Columbia University, New York, NY 10027, USA}
\email{rrdiesing@ias.edu}

\author[0000-0003-4690-2774]{Ludwig M. B{\"o}ss}
\affiliation{Department of Astronomy and Astrophysics, The University of Chicago, Chicago, IL 60637, USA}
\email{lboess@uchicago.edu}

\author[0000-0003-0939-8775]{Damiano Caprioli}
\affiliation{Department of Astronomy and Astrophysics, The University of Chicago, Chicago, IL 60637, USA}
\affiliation{Enrico Fermi Institute, The University of Chicago, Chicago, IL 60637, USA}
\email{caprioli@uchicago.edu}

\section{Introduction} \label{sec:intro}

Interpreting observations of the non-thermal emission from galaxy clusters requires a complete picture of particle acceleration and transport in the intracluster medium (ICM). In particular, merging galaxy clusters often exhibit radio halos and relics--diffuse, synchrotron-emitting structures with no galaxy counterparts--implying efficient particle acceleration \citep[see, e.g.,][for relevant reviews]{brunetti+14, vanWeeren+19}. A natural theoretical explanation invokes merger-driven shocks \cite[e.g.,][]{miniati+01,  ryu+03, blasi+07b,  kang+12, pinzke+13, ha+18b, botteon+20, wittor+20, smolinski+23}, which accelerate (or, in some cases, reaccelerate) charged particles (cosmic rays, CRs) via diffusive shock acceleration (DSA, e.g., \cite{krymskii77, axford+77p, bell78a, blandford+78}; see also \cite{kim+21} for a detailed discussion of electron acceleration at cluster shocks).

In this picture, merger shocks (re)accelerate protons and electrons with power-law distributions up to a maximum energy set by the acceleration time, $\tau_{\rm acc}$, which is proportional to the diffusion timescale, $\tau_{\rm diff}$ \citep[e.g.,][]{drury83, blasi+07}. In the case of protons, which experience negligible energy losses aside from adiabatic expansion, one can approximate the maximum energy by requiring that $\tau_{\rm acc}$ be shorter than the age of the shock, and that particles remain confined to the vicinity of the shock during that time \citep[e.g.,][]{hillas05, bell+13, diesing23}. As such, the maximum proton energy, $E_{\rm max}$, which determines the highest energy $\gamma$-rays a cluster will emit--depends on the nature of CR transport in the ICM. Meanwhile, in the case of electrons, synchrotron losses are significant, and one can approximate the maximum electron energy by equating $\tau_{\rm acc} \sim \tau_{\rm diff}$ with the loss timescale.

Recently, \cite{reichherzer+25} put forward a new picture of CR transport in high-$\beta$ plasmas (i.e., environments, such as the ICM, in which the thermal pressure greatly exceeds the magnetic pressure). 
In these environments, pressure anisotropies can trigger the firehose and mirror instabilities, which distort the magnetic field on small (plasma) scales. 
Very interestingly, these fluctuations seem  to generally saturate at the level of $\delta B/B_0\sim 0.3$ at scales of a few ion skin depths, regardless of the amplitude of the anisotropy and the plasma $\beta$, 
hence providing a universal minium level of ICM turbulence \citep[][]{kunz+14b, reichherzer+25}. 

Such magnetic structures, or ``micromirrors," can efficiently scatter CRs and control their transport. The end result can be described in terms of an effective diffusion coefficient that pervades the galaxy cluster, and which does not require CR streaming to excite magnetic waves \citep[e.g.,][]{skilling75a,bell04,amato+09, zweibel13, bykov+13}. 
Of course, such instabilities may still be triggered, especially in regions of CR over-densities, thereby confining CRs even further. Thus, micromirrors enforce a \emph{maximum} allowed diffusion coefficient and, by extension, a \emph{minimum} value of $E_{\rm max}$.

In this paper we investigate the implications of micromirror confinement for shock acceleration in the ICM, considering their role in both merger and accretion shocks. 
More specifically, we estimate the minimum value of $E_{\rm max}$ allowed in an ICM environment, and discuss micromirrors in the context of large-scale cosmological simulations and $\gamma$-ray observations. While radio observations provide evidence for electron acceleration in the ICM, we choose to largely focus on proton acceleration in this work, as non-thermal ions are far more likely to be dynamically important than non-thermal electrons. Moreover, since ion maximum energies are not loss-limited, the implications of micromirror confinement are more significant for these species.
It is also worth noting that radio halo emission may be the result of relativistic electrons being reaccelerated via magnetic turbulence \cite[e.g.,][]{brunetti+05, brunetti+07, blasi+07b, brunetti+17, vanweeren+17}, which is not directly related to spatial diffusion.

This paper is organized as follows. We present a series of analytic estimates in Section \ref{sec:analytic}, including an estimate of the  micromirror-enforced maximum energy, $E_{\rm max}$, in Section \ref{subsec:emax}, a discussion of CR transport in the context of cosmological simulations in Section \ref{subsec:transport}, and a discussion of observational constraints in Section \ref{subsec:obs1}. In Section \ref{sec:sims}, we present cosmological simulations of a Coma-like cluster, including a CR treatment consistent with the estimates presented in Section \ref{sec:analytic} (Section \ref{subsec:setup}), and compare our simulated cluster to $\gamma$-ray observations (Section \ref{subsec:obs2}). We summarize in Section \ref{sec:discussion}.

\section{Analytic Estimates}
\label{sec:analytic}

To better understand the implications of CR confinement in the ICM, herein we perform a series of simple estimates pertaining to shock acceleration in an environment where micro mirrors control CR transport. Throughout this section, we invoke the following assumptions and definitions:

\begin{itemize}
    \item The ICM hosts an ensemble of mostly weak shocks (with Mach number $ M\gtrsim 1$) due to a combination of merger history and accretion. These shocks (re)accelerate hadrons and electrons via diffusive shock acceleration (DSA). Typical shock velocities are thus $v_{\rm sh,8} \equiv\vsh/(1000 \rm{\ km \ s^{-1}})\gtrsim 1$.
    \item Typical temperatures, magnetic field strengths, and ambient number densities are of the order $T_{\rm 5 keV} \equiv T/(\rm 5 \ keV) \sim 1$, $B_{3\mu G} \equiv B/(3\mu$G), $\sim 1$, and $n_{\rm ICM, -3} \equiv n_{\rm ICM}/(10^{-3} \rm{ \ cm^{-3}})\sim1$ \citep{kunz+24, govoni+17}.
    \item Particle transport can be approximated as diffusive, with an effective diffusion coefficient, $\kappa_{\rm eff}$, set by micromirror confinement as described in \cite{reichherzer+25}.
    
\end{itemize}

\subsection{The maximum proton energy} \label{subsec:emax}

In the context of shock acceleration, the most important implication of micromirror confinement is a lower limit on the maximum proton energy, $E_{\rm max}$. Namely, as we will demonstrate, micromirrors ensure that typical clusters produce, at minimum, $\sim 100$ GeV particles.

To estimate $E_{\rm max}$, we start with the micromirror diffusion coefficient calculated in \cite{reichherzer+25} (Equation 5),
\begin{equation}
\begin{split}
    \kappa_{\rm mm} \simeq  10^{24} \text{ cm$^2$ s$^{-1}$} \\ \times \ Z^{-2}T_{\rm 5 keV} B_{\rm 3\mu G}^{-1}\delta B_{\rm mm, 1/3}^{-2}E_{\rm GeV}^2 .
\end{split}
\end{equation}
Here, $Z$ is the CR atomic number, $\delta B_{\rm mm,1/3} \equiv 3\delta B/B_0$ is the micromirror amplitude, taken to be of order $1/3$ as in \cite{reichherzer+25}, and $E_{\rm GeV}$ is the particle energy normalized to 1 GeV. Note that $\kappa_{\rm mm}$ is sensitive to the characteristic micromirror scale, which depends on the thermal ion gyroradius and thus the temperature \citep{rincon+15}.

Assuming particles are accelerated via DSA, this diffusion coefficient gives an acceleration time, $\tau_{\rm acc} \simeq 6R/(R-1)\kappa_{\rm mm}/\vsh^2$ \citep[][]{drury83},
\begin{equation}
\begin{split}
    \tau_{\rm acc} \simeq 19 \text{ yr} \times R/(R-1) \\ \times \  Z^{-2}  T_{\rm 5 keV}B_{\rm 3\mu G}^{-1} \delta B_{\rm mm, 1/3}^{-2}E_{\rm GeV}^2v_{\rm sh, 8}^{-2} 
\end{split}
\end{equation}
where $R \equiv \rho_2/\rho_1 = (\gamma_{\rm ad}+1)M^2/((\gamma_{\rm ad}-1)M^2+2)$ is the compression ratio, and $\gamma_{\rm ad} \simeq 5/3$ is the adiabatic index of the gas. 

Equivalently, this gives a diffusion length, $\ell_{\rm diff} = \kappa_{\rm mm}/\vsh$,
\begin{equation}
\begin{split} 
   \ell_{\rm diff} \simeq 3\times10^{-3} \text{ pc } \\ \times \ Z^{-2}  \ T_{\rm 5 keV}B_{\rm 3\mu G}^{-1}  \delta B_{\rm mm, 1/3}^{-2}E_{\rm GeV}^2v_{\rm sh, 8}^{-1} \rm \ .
\end{split}
\end{equation}
Clearly, for typical parameters, GeV particles are easily accelerated, since $\tau_{\rm acc} \ll \tau_{\rm sh}$ and $\ell_{\rm diff} \ll \ell_{\rm sh}$, where $\tau_{\rm sh}$ and $\ell_{\rm diff}$ are characteristic time and length scales of a typical radio relic.

However, it is important to note that $\kappa_{\rm mm}$ is a \emph{minimum} value of the diffusion coefficient if micromirrors do not fill the entire ICM volume. As such, \cite{reichherzer+25} obtains a steady-state \emph{effective} diffusion coefficient, $\kappa_{\rm eff}$, given by
\begin{equation}
    \kappa_{\rm eff} \simeq \frac{\kappa_{\rm mm}}{f_{\rm mm}+(1-f_{\rm mm})\kappa_{\rm mm}/\kappa_{\rm max}}.
\end{equation}
Here, $f_{\rm mm}$ is the micromirror filling fraction (estimated to be $\simeq 0.1$ in kinetic simulations) and $\kappa_{\rm max}$ is the diffusion coefficient in the remaining volume of the ICM. Assuming $\kappa_{\rm mm} \ll \kappa_{\rm max}$, we obtain $\kappa_{\rm eff} \simeq \kappa_{\rm mm}/f_{\rm mm}$. Defining $f_{\rm mm,-1} \equiv f_{\rm mm}/10^{-1}$, our estimates for $\tau_{\rm acc}$ and $\ell_{\rm diff}$ become, 
\begin{equation}
\label{eq:tacc2}
\begin{split}
    \tau_{\rm acc} \simeq 190 \text{ yr } \times R/(R-1) \\ \times Z^{-2}T_{\rm 5 keV}B_{\rm 3\mu G}^{-1}\delta B_{\rm mm, 1/3}^{-2} E_{\rm GeV}^2v_{\rm sh, 8}^{-2}f_{\rm mm, -1}^{-1},
\end{split}
\end{equation}
\begin{equation}
\begin{split}
   \ell_{\rm diff} \simeq 3\times10^{-2} \text{ pc } \\ \times \ Z^{-2}T_{\rm 5 keV}B_{\rm 3\mu G}^{-1} \delta B_{\rm mm, 1/3}^{-2} E_{\rm GeV}^2v_{\rm sh, 8}^{-1}f_{\rm mm, -1}^{-1}.
\end{split}
\end{equation}
These values for $\tau_{\rm acc}$ and $\ell_{\rm diff}$ are still extremely small compared to typical merger shock lifetimes and length scales, $\tau_{\rm sh}$ and $\ell_{\rm sh}$ \citep[e.g.,][]{brunetti+14}. We also emphasize that, by assuming that transport between micromirrors is effectively ballistic (recall that we approximate $\kappa_{\rm mm}/\kappa_{\rm max} =0$), we neglect any confinement other than that of micromirrors. This ensures that our estimates are relatively conservative.

Let us now explicitly estimate the maximum proton energy, assuming $E_{\rm max}$ is set by equating $\tau_{\rm acc} = \tau_{\rm sh}$; for typical merger and accretion shocks, the acceleration time presents a more stringent limit than the diffusion length. Rewriting Equation \ref{eq:tacc2}, we obtain
\begin{equation}
\begin{split}
    E_{\rm max, \ mm} \simeq 73 \text{ GeV } \times (R/(R-1))^{-1/2} \\ \times \ Z\tau_{\rm sh, 6}^{1/2} T_{\rm 5keV}^{1/4}B_{\rm 3\mu G}^{1/2}\delta B_{\rm mm, 1/3}v_{\rm sh, 8}f_{\rm mm, -1}^{1/2},
\end{split}
    \label{eq:emax_mm}
\end{equation}
where $\tau_{\rm sh,6}$ is the shock age normalized to 1 Myr. Consider a shock with $M = 2$ ($R \simeq 2.3$). The speed of sound in a gas with $k_{\rm B}T=5\rm{\ keV}$ and mean molecular weight $\mu=0.6$ is roughly $1200\rm{ \ km \ s^{-1}}$. Thus, we obtain $\vsh \simeq 2400\rm{ \ km \ s^{-1}}$ and $E_{\rm max} \simeq 130 \rm \ GeV$ on a Myr timescale. 

For comparison, let us also consider a diffusion coefficient comparable to the galactic one: $\kappa_{\rm gal} \simeq 3\times10^{28}(E/\rm GeV)^{\delta} \ cm^2 \ s^{-1}$ \citep[e.g.,][]{strong-ptuskin+07}. We now obtain
\begin{equation}
    \tau_{\rm acc} \simeq 5.7\times10^5 \text{ yr } \times R/(R-1)E_{\rm GeV}^{\delta}v_{\rm sh, 8}^{-2},
\end{equation}
\begin{equation}
    E_{\rm max, \ gal} \simeq 1 \text { GeV } \times [1.75 \times (R/(R-1))^{-1}\tau_{\rm sh, 6}v_{\rm sh, 8}^{2}]^{1/\delta} .
\end{equation}
Assuming a value consistent with that inferred for the Galaxy, $\delta = 1/3$, we obtain $E_{\rm max} \simeq 180$ GeV. Note that this value is only slightly higher than the one we estimated with micromirror confinement. In short, micromirrors enforce a maximum proton energy that would otherwise require particle confinement comparable to that in the Milky Way.

Thus far, we have focused on parameters that are broadly consistent with merger shocks. However, accretion shocks can be meaningfully faster and longer-lived, and are therefore capable of accelerating particles to even higher energies. Consider a relatively fast accretion shock with typical parameters $\tau_{\rm sh} = 1$ Gyr, $B_0 = 10^{-2}\mu$G, $\vsh = 5000 \rm \ km \ s^{-1}$, and $M\gg 1$ (i.e., $R\simeq4)$ \citep[e.g.][]{carretti+22, ha+23}. 
In this case, we expect $E_{\rm max} \simeq 580$ GeV. 
Thus, we predict that while accretion shocks can indeed accelerate particles to higher energies, they are unlikely to produce--with micromirrors alone--the 1-10 TeV particles required to be detectable by very-high-energy (VHE) $\gamma$-ray telescopes such as MAGIC, H.E.S.S., or CTA. 

However, it is important to recall that the $E_{\rm max}$ calculated here is a conservative estimate. In practice, micromirrors also create complex magnetic field structures in the ICM, thereby ensuring that an appreciable fraction of any shock surface is quasi-parallel (i.e., the shock normal aligns with the ambient field), and thus favorable to particle injection \citep[e.g.,][]{caprioli+14a}. As such, micromirrors can lead to efficient particle acceleration, which in turn leads to magnetic field amplification \citep[e.g.,][]{bell78a,bell04,amato+09,bykov+13}, further suppressing the CR diffusion coefficient \citep{reville+13,caprioli+14b,caprioli+14c}.  
For example, if CR streaming instabilities are sufficient to generate resonant fluctuations with $\delta B/B\sim1$, we can instead estimate $E_{\rm max}$ in the Bohm limit (recall that $\kappa_{\rm Bohm} = cr_{\rm L}/3$, where $r_{\rm L}$ is the Larmor radius):
\begin{equation}
\begin{split}
E_{\rm max, \ Bohm} \simeq 5.3\times10^{6}\text{ GeV }\times (R/(R-1))^{-1} \\ Z\tau_{\rm sh,6} B_{\rm 3\mu G}v_{\rm sh,8}^2. 
\end{split}
\end{equation}
This estimate--which, notably, does not include any amplification past $\delta B/B \sim 1$--predicts that our benchmark merger shock will accelerate 17 PeV protons. As such, a search for $\gamma$-ray emission from galaxy clusters with CTAO 

\subsection{Cosmic ray transport}
\label{subsec:transport}

Of course, micromirrors also have strong implications for CR transport in the ICM; extensive discussion of these implications can be found in \cite{reichherzer+25}. Here we add a brief note about CR transport in cosmological simulations. Generally speaking, these simulations, when they include CRs, approximate CR transport as advective \citep[e.g.,][]{boess+23b, vazza+25}, as this assumption is the most computationally inexpensive. As we will show, micromirror confinement ensures that this assumption is theoretically motivated. Namely, the advection time is given by,
\begin{equation}
    \tau_{\rm adv} \sim R/\vsh \simeq 1 \text{ Gyr }\times R_{\rm Mpc}v_{\rm sh,8}^{-1},
\end{equation}
where $R_{\rm Mpc} \equiv R/(1 \ \rm Mp)$ is the size of the galaxy cluster. During this time, CRs diffuse over a distance $\lambda_{\rm diff} = \sqrt{6\kappa_{\rm eff}\tau_{\rm adv}}$ and we approximate,
\begin{equation}
\begin{split}
     \lambda_{\rm diff}\simeq 440 \times\text{ pc } \times Z^{-1} \\ \times \  T_{\rm 5keV}^{-1/4}B_{\rm 3\mu G}^{-1/2}\delta B_{\rm mm, 1/3}^{-1}E_{\rm GeV}R_{\rm Mpc}^{1/2}v_{\rm sh,8}^{-1/2}f_{\rm mm, -1}^{-1/2}.
\end{split}
\end{equation}
Thus, $\lambda_{\rm diff} \ll R$ and we can safely approximate CR transport as advective. As such, we will transport CRs via advection in the cosmological simulations presented in Section \ref{sec:sims}.

\subsection{Observational limits} \label{subsec:obs1}

\begin{figure}[ht]
    \centering
    \includegraphics[width=\linewidth, clip=true,trim= 15 0 40 25 ]{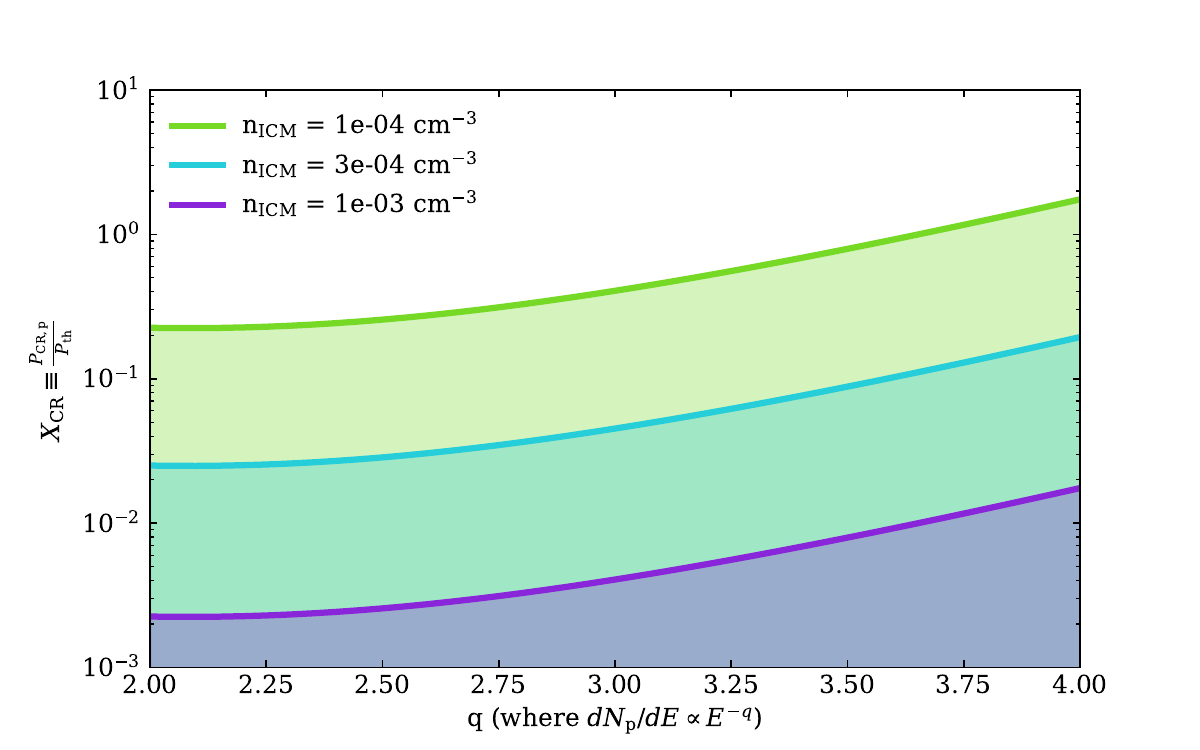}
    \caption{Allowed ratios of the maximum average CR pressure to the average ICM thermal pressure ($X_{\rm CR}$; shaded regions) as a function of the proton power-law slope, $q$, in the Coma cluster, based on Fermi-LAT upper limits \citep{ackermann+16}. Colors correspond to different ambient densities, $n_{\rm ICM}$. To remain consistent with observations $X_{\rm CR}$ needs to be fairly small ($\lesssim 10^{-2}-10^{-3}$), depending on the average ambient density.}
    \label{fig:pressureratio}
\end{figure}

\begin{figure*}[ht]
    \centering
    \includegraphics[width=\linewidth]{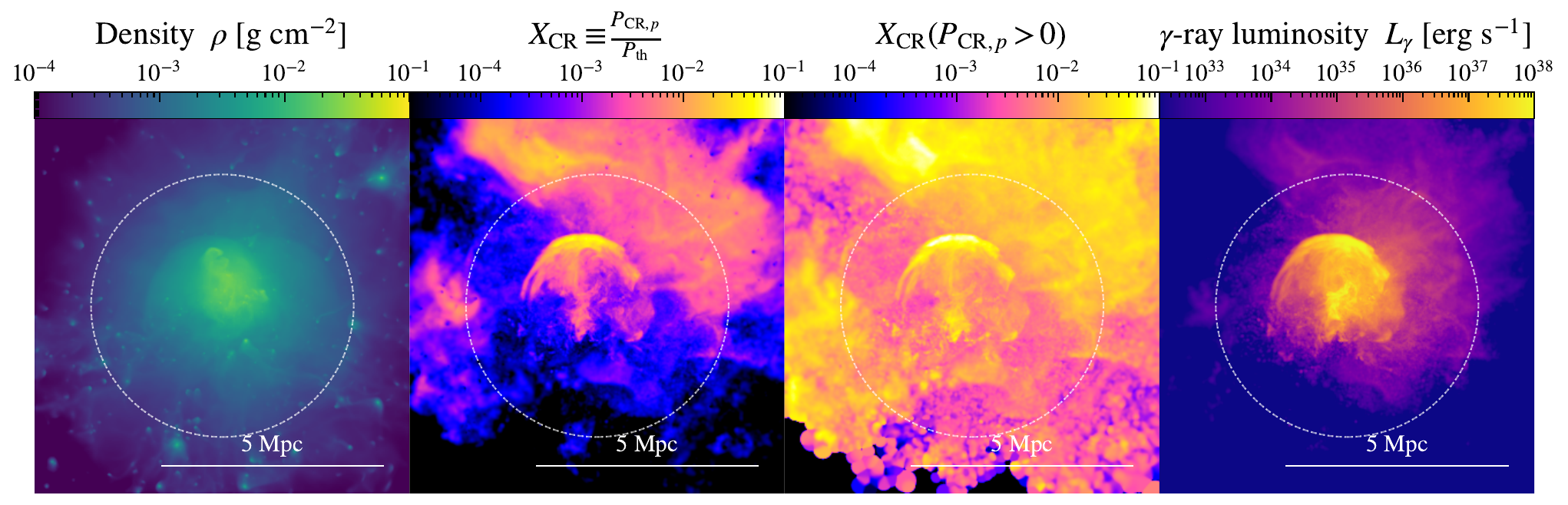}
    \caption{Density (left panel), CR to thermal pressure ratio (center panels), and $\gamma$-ray luminosity (right panel) maps of our fiducial cluster simulation, a Coma-like cluster with a viral mass of $M_\mathrm{vir} \approx 2 \times 10^{15} M_\odot$. The circle illustrates $r_{200}$ of the cluster.\\}
    \label{fig:maps}
\end{figure*}

Of course, if galaxy cluster shocks are readily capable of accelerating $\sim 100$ GeV particles, it is worth understanding the ICM conditions required to ensure that the resulting CR population does not produce $\gamma$-ray emission in conflict with observations. Namely, accelerated protons produce $\gamma$-ray emission via proton-proton collisions, which produce neutral pions that decay into $\gamma$-rays. However, at present, there is no definitive detection of $\gamma$-rays from galaxy clusters \cite{ackermann+14}. While possible detections in the direction of the Coma cluster have been reported in, e.g., \cite{keshet+17, keshet+18, adam+21}, source confusion is likely. As such, we instead look to the upper limits presented in \cite{ackermann+16}, which again focuses on the Coma cluster. Note that, due to its proximity and recent merger history \citep[e.g.][]{burns+1994}, Coma provides some of the strongest available constraints on hadronic acceleration. 

Herein we introduce a simple estimate of the maximum CR to thermal pressure ratio, $X_{\rm CR} \equiv P_{\rm CR,p}/P_{\rm th}$ allowed by \cite{ackermann+16}. While works such as \cite{brunetti+17, ha+20, wittor+20} provide more detailed constraints on hadron acceleration \citep[see also][for a detailed review]{wittor+21b}, we develop this simple approximation in order to set a benchmark against which to compare our cluster simulation presented in Section \ref{sec:sims}. 

Based on Figure 3 in \cite{ackermann+16}, we can estimate that the hadronic $\gamma$-ray flux at GeV energies cannot exceed $\sim 1.6\times10^{-13} \rm{\ erg \ cm^{-2} \ {s^{-1}}}$. The hadronic $\gamma$-ray flux at energy $E$ can be approximated as,
\begin{equation}
    \Phi_\gamma(E) \simeq 
    \frac{E^2n_{\rm ICM}\sigma_{\rm pp}}{4\pi d^2}\frac{dN_{\rm p}(E/\chi)}{dE},
\end{equation}
where $\sigma_{\rm pp} \simeq 31$ mb is the proton-proton cross section, $d$ is the distance to the Coma cluster, taken to be $\simeq 100$ Mpc \citep[as in][]{ackermann+16}, $dN_{\rm p}(E)/dE$ is the proton distribution, and $\chi \equiv E_\gamma/E_{\rm p} \simeq 0.1$. 
Assuming the CR proton distribution takes a power-law form,
\begin{equation}
    \frac{dN_{\rm p}(E)}{dE} = AE_{\rm GeV}^{-q}
\end{equation}
we obtain the constraint,
\begin{equation}
    A \lesssim 8\times10^{67} \text{ erg$^{-1}$ } \kappa^{-q} n_{\rm ICM,-3}^{-1}d_{\rm 100}^2,
\end{equation}
which corresponds to an average CR energy density,
\begin{equation}
\begin{split}
    \epsilon_{\rm CR} \lesssim 10^{-1} \text{ eV cm$^{-3}$} \\ \times \ \chi^{-q}(2-q) (\gamma_{\rm max}^{2-q}-\gamma_{\rm min}^{2-q})n_{\rm ICM,-3}^{-1}d_{100}^2R_{\rm 2 Mpc}^{-3} .
\end{split}
\end{equation}
Here, $\gamma_{\rm min}$ and $\gamma_{\rm max}$ refer to the minimum and maximum CR Lorentz factor, respectively.

Figure \ref{fig:pressureratio} shows $P_{\rm CR} = \epsilon_{\rm CR}/3$ relative to the thermal pressure, $P_{\rm ICM} \simeq n_{\rm ICM}k_{\rm B}T_{\rm ICM}$, assuming $k_{\rm B}T_{\rm ICM} = 8.3 \rm{ \ keV}$ \citep{wik+11}, $\gamma_{\rm min} = 4k_BT_{\rm ICM}/(m_{\rm p}c^2) + 1$, and $\gamma_{\rm max} = 10$. A correction to the form of $dN_{\rm p}(E)/dE$--namely, a factor of $\beta \equiv v_{\rm p}/c$--is also included to ensure power-law behavior in momentum (this becomes important when $q \gg 2$).

Note that, based on the results shown in Figure \ref{fig:pressureratio}, we do not expect $\gamma$-rays from accretion shocks to be important; for a typical accretion shock ambient density of $n_{\rm ICM} \simeq 10^{-5}$ in the warm-hot intergalactic medium (WHIM) \citep[][]{dave+01}, the GeV flux only approaches the Fermi-LAT sensitivity if $X_{\rm CR} \gg 1$. 

\section{Simulations}
\label{sec:sims}
To test our analytic approximations--in particular, the observational limits presented in Section \ref{subsec:obs1}--we conduct MHD simulations of a massive galaxy cluster, including a prescription for the maximum proton energy, $\Emax$, that is consistent with a diffusion coefficient set by micromirror confinement (see Section \ref{subsec:emax} and Equation \ref{eq:emax_mm}). Furthermore, consistent with the calculations presented in Section \ref{subsec:transport}, we transport CRs via advection only.

\subsection{Setup}
\label{subsec:setup}

We performed zoom-in simulations of a galaxy cluster with a mass of $M_\mathrm{vir} \approx 2 \times 10^{15} M_\odot$, undergoing a merger at $z=0$.
The initial conditions for this setup are constructed by spatial hyper-refinement of a cosmological volume simulation \citep[see][for details]{bonafede+11} to a target resolution of $m_\mathrm{DM} = 4.7 \times 10^7 M_\odot$ and $m_\mathrm{gas} = 8.7 \times 10^6 M_\odot$.
This allows us to simulate a high-resolution galaxy cluster in a fully cosmological context.
The initial conditions for this cluster, albeit at lower resolution, were found to be a good match for a Coma analog in \citet{bonafede+11}
 and therefore lends itself nicely for comparisons to observations.
 The same initial conditions at different resolutions have been used to study magnetic field amplification \citep[][]{steinwandel+22, steinwandel+24}, shocks \citep[][]{zhang+20c} and radio relics \citep[][]{boess+23b}.
 
 The simulation was performed with \textsc{OpenGadget3} \citep[][]{groth+23}, a cosmological Tree-SPH code based on \textsc{Gadget2} \citep[][]{springel05}.
We employ a modern SPH scheme with higher order kernels \citep[][]{beck+16a}, an on-the-fly shock finder \citep[][]{beck+16b}, non-ideal MHD in the form of magnetic diffusion and dissipation \citep[][]{dolag+09b, bonafede+11}, a modern hyperbolic divergence cleaning scheme \citep[][]{tricco+16} and an on-the-fly spectral CR model \citep[][]{boess+23a}.

For the present work, we are interested in the CR proton component of the simulation.
The spectral CR model solves the diffusion-advection equation in the advective limit, without accounting for CR diffusion/streaming, in line with the assumption of efficient micromirror confinement.
We account for CR acceleration at shocks and adiabatic changes, neglecting Coulomb losses for protons, since the cooling time for CR protons in typical densities of galaxy clusters is of the order of the Hubble time \citep[e.g.,][]{berezinsky+97,brunetti+14}.
For this work, we used a constant CR acceleration efficiency of $\eta = 0.1$ for shocks with a sonic Mach number $\mathcal{M}_s \geq 2$, but without explicit dependence on shock obliquity \citep{kang+13b, caprioli+14a, wittor+20}. 

CRs are injected into the momentum range $p \in [p_\mathrm{inj}, p_\mathrm{max}]$, where $p_\mathrm{inj} = 3.5 \times \sqrt{2 k_B m_p T_2}$ with $T_2$ being the downstream temperature and $p_\mathrm{max} = E_\mathrm{max}/c$, using $E_\mathrm{max}$ in Eq. \ref{eq:emax_mm}.
$E_\mathrm{max}$ is computed on-the-fly using the local fluid quantities with the simplification of using a constant $\tau_\mathrm{sh,6} = 10$ Myr.
We represent the spectrum with 8 bins in the momentum range $p \in [0.1, 10^3] \: m_p c$.
To account for higher maximum momenta than the one arbitrarily chosen at the start of the simulation, we also evolve a spectrum cutoff $p_\mathrm{cut}$ which allows us to extend the initial momentum range beyond $p = 10^3 \: m_p c$ to account for larger $p_\mathrm{max}$.
Our solver requires at least two filled bins to evolve a spectrum, hence we made the conservative choice to not inject CRs if Eq.~\ref{eq:emax_mm} yields $E_\mathrm{max} < 10$ GeV.

The expected $\gamma$-ray emission of the cluster can be computed directly from the simulated proton spectra according to the description in \citet{boess+25}.

\begin{figure}
    \centering
    \includegraphics[width=\linewidth]{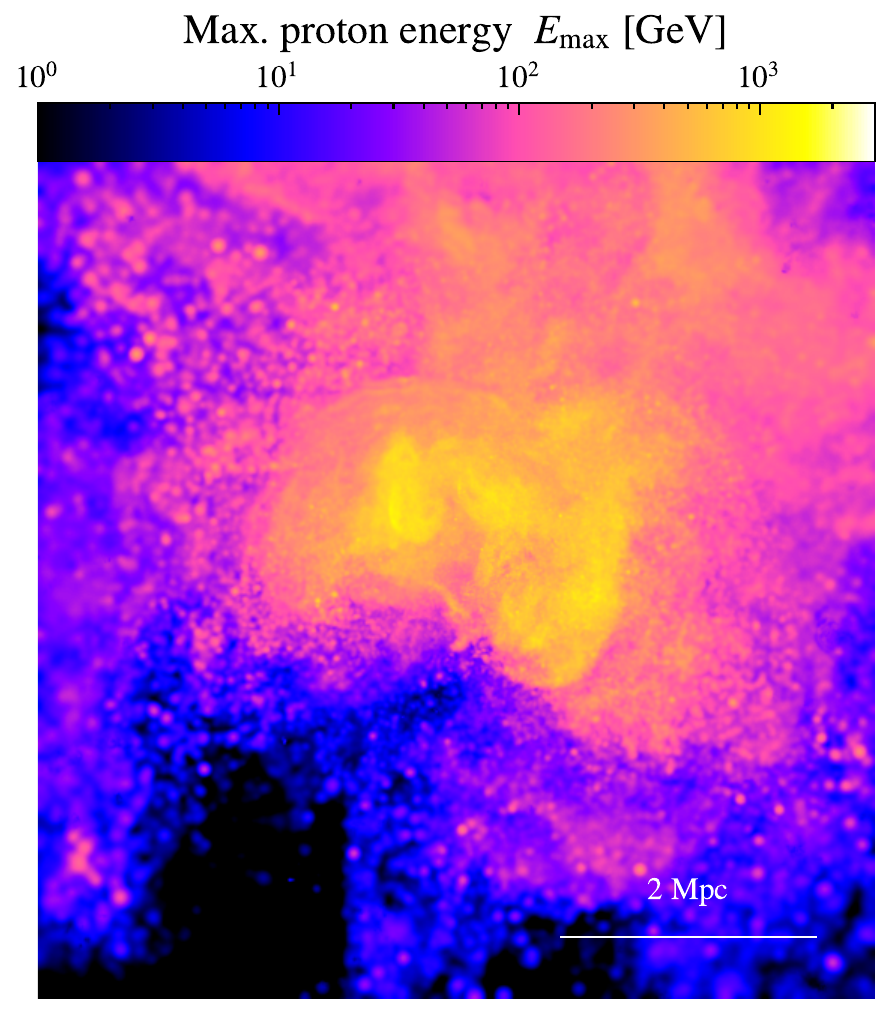}
    \caption{Maximum energy of the shock accelerated protons, following Equation~\ref{eq:emax_mm}.}
    \label{fig:g55_pmax}
\end{figure}

\begin{figure}
    \centering
    \includegraphics[width=\linewidth]{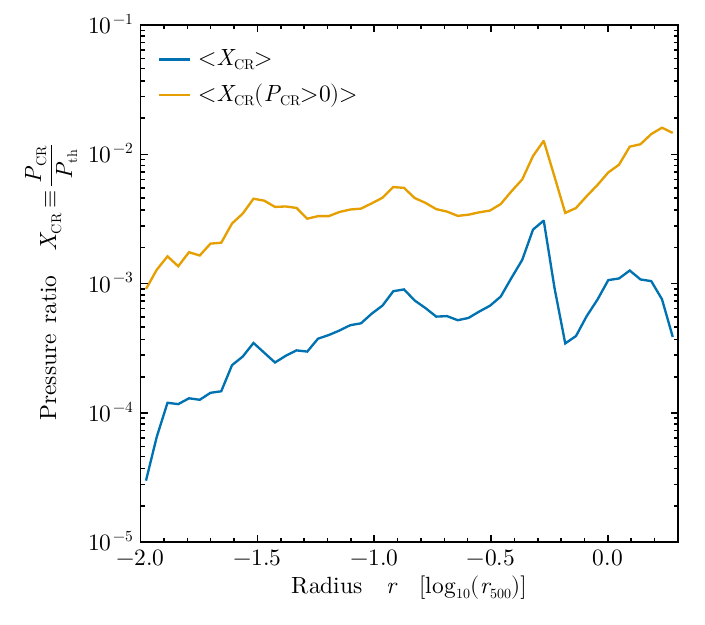}
    \caption{Average CR to thermal pressure ratio, $X_{\rm CR}$, as a function of radius for a simulated Coma analog. The blue line shows $X_{\rm CR}$ averaged over the entire shell at radius $r$, while the orange line only includes regions where the CR pressure ($P_{\rm CR}$) is nonzero. In both cases, $X_{\rm CR}$ generally falls well below the maximum allowed values illustrated in Figure \ref{fig:pressureratio}.}
    \label{fig:profiles}
\end{figure}

\subsection{Results}

Figure \ref{fig:maps} shows the simulated cluster at $z=0$.
From left to right we show surface density, CR to thermal pressure ratio $X_\mathrm{CR}$ (center two panels) and $\gamma$-ray luminosity.
For $X_\mathrm{CR}$ we show the mean value for $X_\mathrm{CR}$ calculated along the whole line of sight both all SPH particles (center left) and only for SPH particles that contain CR population (center right);
this illustrates that, while we do reach values of $X_\mathrm{CR} \sim 10\%$ in shocked regions, the overall energy density of CR protons is still small.
Similarly, we can see that the $\gamma$-ray luminosity is confined only to the inner-most region, where active shocks are accelerating CRs in the highest-density region of the cluster.

Figure~\ref{fig:g55_pmax} shows the maximum energy of shock-accelerated protons calculated via Equation \ref{eq:emax_mm}.
In general, CR protons at merger shocks are accelerated up to $E_\mathrm{max} \sim $ TeV; when advected downstream, they suffer adiabatic losses and their maximum energy decreases.
As a result, the majority of the cluster volume is filled with CR protons with $E_\mathrm{max} \sim 100-300$ GeV.

\subsection{Observational implications}
\label{subsec:obs2}
As discussed in previous sections, micromirrors enforce a minimum value of $E_{\rm max}$ accelerated by shocks in the ICM. 
As such, we expect ICM shocks to produce CRs at energies that are readily detectable by current instruments (in particular, Fermi-LAT). However, the actual flux of these CRs is set by the hydrodynamics of the shocks themselves (shock velocity, ambient density), as well as the proton acceleration efficiency (constrained to be $\sim0.1$ for acceleration from the thermal pool by a quasi-parallel shock \citep[e.g.,][]{caprioli+14a}). 
Thus, even with $E_{\rm max}$ set by micromirror confinement, we find that the GeV $\gamma-$ray flux accelerated by our Coma analog is insufficient to be detected by Fermi-LAT. 

To illustrate this point, we show the CR to thermal pressure ratio, $X_{\rm CR}$ as a function of radius in Figure \ref{fig:profiles}. 
Typical values of $X_{\rm CR}$ hover between $10^{-4}$ and $10^{-3}$ which, as our simple calculations in Section \ref{subsec:obs1} illustrate (see, in particular, Figure \ref{fig:pressureratio}), are likely too low to be detectable with current instruments. 

\begin{figure}[ht]
    \centering
    \includegraphics[width=\linewidth]{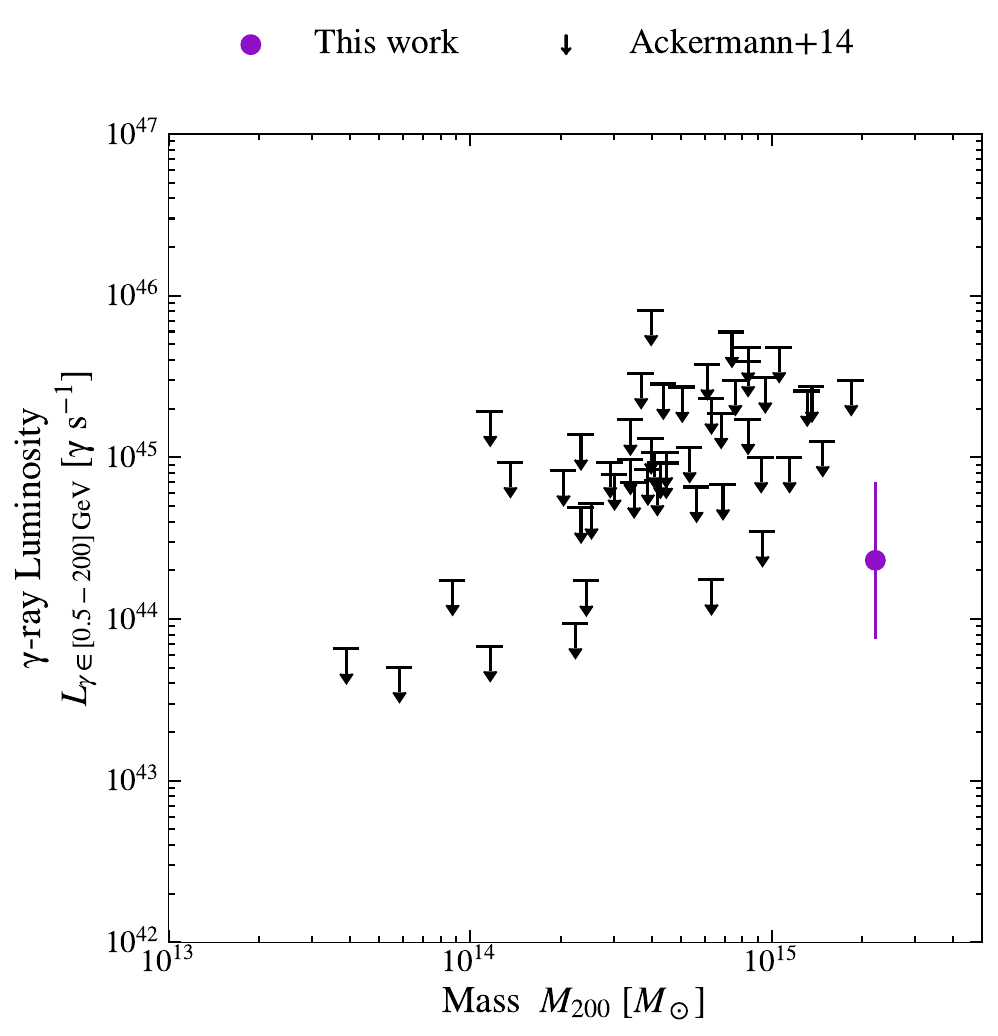}
    \caption{Integrated $\gamma$-ray luminosity of our simulated Coma analog (purple point), compared to upper-limits from a sample of clusters measured with Fermi-LAT \citep{ackermann+14}. The errorbar indicates the maximum impact of the ultra-relativistic approximation on our results. Our modeled cluster falls well below these upper-limits, implying that, while micromirror confinement can enhance the maximum proton energy, it does not yield a tension with observations. }
    \label{fig:upperlims}
\end{figure}

This result is consistent with the non-detection of $\gamma$-rays from galaxy clusters (with the possible exception of Coma, though in this case source confusion is likely \citep{adam+21}). More specifically, Figure \ref{fig:upperlims} directly compares our Coma analog to Fermi-LAT observations of clusters measured in \cite{ackermann+14}; our modeled cluster falls well below their measured upper limits.
We note, however, that the ultra-relativistic approximation used in our modeling of CR injection can propagate to our results for $\gamma$-ray emission.
For weak shocks with $\mathcal{M}_s = 2$, the lower limit of our injection, the error introduced by using the ultra-relativistic approximation for computing the injection norm at $p_\mathrm{inj}(10^8 \: \mathrm{K}) \approx 10^{-2} \: m_p c$, leads to an over-estimation of the $\gamma$-ray emissivity of a factor of five and at $\mathcal{M}_s = 2.3$, the lower limit for acceleration found in \citet{ryu+19} to a factor of two.
For $\mathcal{M}_s = 3$ shocks, the difference is negligible, however and for stronger shocks, we under-estimate the emissivity by less than 10 per cent.
Other effects, like the dependence of acceleration efficiency on shock obliquity, and the model for sonic Mach number dependent acceleration efficiency can have significantly stronger effects \citep[see discussion in][]{boess+25}.
Nonetheless, we indicate the range a factor of three error introduces on out results in Fig.~\ref{fig:upperlims}.

In principle CR electrons could also contribute to the total $\gamma$-ray emission from clusters via inverse Compton (IC) up-scattering of cosmic microwave background (CMB) photons. However, to achieve the $\sim$ GeV photon energies required to be detectable by current $\gamma$-ray instruments (namely, Fermi-LAT), one requires electrons with energies $\sim 540$ GeV (assuming 160 GHz CMB photons). Thus, assuming micromirror confinement, only accretion shocks are capable of achieving such energies. However, electrons will also suffer substantial IC losses on the timescale required to accelerate to $\sim 100$ GeV energies. Equating the acceleration timescale calculated in Equation \ref{eq:tacc2} with the IC loss timescale for ultra-relativistic electrons, $\tau_{\rm IC} = 3m_{\rm e}^2c^3/(4\sigma_{\rm T}u_{\rm rad}E)$, where $\sigma_{\rm T}$ is the Thomson cross section and $u_{\rm rad}\simeq 0.26$ eV cm$^{-3}$ is the energy density of the CMB, we estimate,
\begin{equation}
\begin{split}
    E_{\rm max, e} = 190 \text{ GeV }\times (R/(R-1))^{-1/3} \\ \times T_{\rm 5keV}^{1/3}B^{1/3}_{\rm 3\mu G}\delta B_{\rm mm,1/3}^{2/3}v_{\rm sh,8}^{2/3}f_{\rm mm,-1}^{1/3}
\end{split}
\end{equation}
Thus, for our benchmark accretion shock ($B_0 = 10^{-2}\mu$G, $\vsh = 5000 \rm \ km \ s^{-1}$, $R\simeq4$), we obtain $E_{\rm max,e}\simeq 73$ GeV, which corresponds to IC photons well outside of Fermi-LAT's sensitivity.

\section{Summary} 
\label{sec:discussion}
Using a combination of analytic estimates and cosmological simulations, we investigated how micromirror confinement in the ICM affects the maximum momentum that accelerated particles can achieve at merger and accretion shocks. We summarize our results below.

\begin{enumerate}
    \item Because micromirrors pervade the ICM, they enforce a baseline value of the maximum ion energy that a merger or accretion shock can (re)accelerate. We emphasize that this value is, in effect, a minimum, since CR-driven instabilities not considered here would provide additional scatter. For protons in a typical ICM, we estimate this baseline $E_{\rm max} \simeq 130$ GeV for a merger shock. Note that this value is comparable to the $E_{\rm max}$ achieved if ICM diffusion were comparable to the Galactic one.
    \item In the presence of micromirrors, accretion shocks can accelerate particles to even higher energies than merger shocks, with $E_{\rm max}\simeq 580 $ GeV. As such, it is plausible that, with the addition of CR-driven magnetic field amplification \citep[e.g.,][]{skilling75a, bell04}, accretion shocks could produce TeV $\gamma$-rays.
    \item Micromirrors ensure that, throughout the ICM, the advection time is much shorter than the diffusion time. As such, cosmological simulations that include CRs can safely invoke the cost-saving assumption that CR transport is purely advective.
    \item In order to remain consistent with $\gamma$-ray upper limits \citep[e.g.,][]{ackermann+14}, we estimate that the average CR to thermal pressure ratio in the Coma cluster, $X_{\rm CR}$, cannot exceed $\sim 10^{-3} - 10^{-2}$, depending on the average ICM density and the slope of the CR spectrum. 
    \item To confirm that shock acceleration in the presence of micromirrors does not produce $\gamma$-ray emission in conflict observations, we performed a cosmological simulation of a Coma-like cluster, including efficient CR acceleration ($\xi_{\rm CR} \equiv P_{\rm CR}/(\rho\vsh^2) = 0.1$) and $E_{\rm max}$ set by micromirror confinement. We find that our simulated cluster exhibits $X_{\rm CR} \sim 10^{-4}-10^{-3}$, yielding a $\gamma$-ray luminosity that is consistent with the results of \cite{ackermann+14}.
\end{enumerate}

\begin{acknowledgements}
We thank Patrick Reichherzer for the helpful discussion. RD gratefully acknowledges support from the Institute for Advanced Study's Fund for Natural Sciences and the Ralph E. and Doris M. Hansmann Member Fund. 
This research was partially supported by NASA grant 80NSSC18K1726 and NSF grants AST-2510951 and AST-2308021 to DC.
The authors acknowledge the Texas Advanced Computing Center (TACC) at The University of Texas at Austin for providing computational resources that have contributed to the research results reported within this paper. URL: http://www.tacc.utexas.edu
\end{acknowledgements}

\bibliographystyle{aasjournalv7}

\end{document}